\documentclass[12pt]{article}
\usepackage{graphicx}
\newcommand{\beq}{\begin{equation}}
\newcommand{\eeq}{\end{equation}}
\newcommand{\bea}{\begin{eqnarray}}
\newcommand{\eea}{\end{eqnarray}}

\begin{document}

\begin{titlepage}
\begin{flushleft}
       \hfill                      {\tt hep-th/0608xxx}\\
       \hfill                       FIT HE - 06-02 \\
\end{flushleft}
\vspace*{3mm}
\begin{center}
{\bf\LARGE Shear viscosity of Yang-Mills theory\\

 \vspace{.2cm}
 in the confinement phase}

\vspace*{5mm} \vspace*{12mm} {\large Iver Brevik \footnote[1]{\tt
iver.h.brevik@ntnu.no} and Kazuo Ghoroku\footnote[2]{\tt
gouroku@dontaku.fit.ac.jp}
}\\
\vspace*{2mm}

\vspace*{2mm}

{\large ${}^{*}$
Department of Energy and Process Engineering, Norwegian University of Science and Technology,
N-7491 Trondheim, Norway}\\
\vspace*{3mm}
{\large ${}^{\dagger}$Fukuoka Institute of Technology, Wajiro,
Higashi-ku}\\
{\large Fukuoka 811-0295, Japan\\}
\vspace*{3mm}

\vspace*{10mm}
\end{center}

\begin{abstract}
In terms of a simple holographic model, we  study the absorption
cross section and the shear viscosity of a pure Yang-Mills field
at low temperature where the system is in the confinement phase.
Then 
we expect that the glueball states are the dominant modes in this
phase. In our holographic model an infrared cutoff $r_m$ is
introduced as a parameter which fixes the lowest mass of the
glueball. As a result the critical temperature of gluon
confinement is estimated to be $T_c\sim 127$ MeV. For $T<T_c$, we
find that both the absorption cross section and  the shear
viscosity are independent of the temperature. Their values are
frozen at the values corresponding to the critical point, for
$0<T< T_c$. We discuss this behavior by considering the glueball
mass and its temperature dependence.

\end{abstract}
\end{titlepage}

\section{Introduction}
Recently, Policastro, Son and Starinets \cite{PSS} have given a
nice application of holographic calculations for the shear
viscosity. This viscosity could be observed in the hot QGP phase
of QCD. The calculation is based on the gravity/gauge
correspondence \cite{Maldacena,GKP,Witten1,report}, and assumes a
strongly coupled 4d YM theory.

With regard to experiments the situation is however unfortunate,
since the quark-gluon plasma one hopes to create in heavy-ion
experiments has relatively low temperature at which the  coupling
constant is strong and the perturbation theory in QCD works
poorly. So we expect that an approach in terms of holographic
models, utilizing the  gravity/gauge correspondence, would be very
useful for the understanding of the various experimental result in
this region.

Recently, we have found that the holographic model available at high
temperature can be extended also to the low temperature quark
confinement phase by introducing an infrared cutoff \cite{GY,GNY}.
This approximate model could give successful results for the
thermal properties, masses of light mesons, decay constants etc,
in the low temperature confinement phase of 4d YM theory. In the
confinement phase, the physical states of QCD would be hadrons.
Then we would be able to see the thermodynamical properties of the
hadron gas by applying our approximate model to the 4d YM theory
in the low temperature phase.

In the present paper we study the shear viscosity, which is
related to the absorption cross section of gravitons \cite{PSS}.
We first study the absorption cross section of gravitons falling
perpendicularly onto D3-branes in the confinement phase. The
calculation is performed in a way similar to that of
Refs.~\cite{Klebanov,DGM}. Then we extend the calculation to the
finite temperature case in order to see the temperature dependence
of the shear viscosity of hadron gas, i.e. the glueball gas in the
present case.

\vspace{.5cm}

\section{Absorption by deformed D3 Brane}
The shear viscosity is related to the absorption cross section, so
we firstly study this quantity at zero temperature to understand
the situation in the confinement phase.

The D3-brane metric can be written as
$$ ds^2= H^{-1/2} \left (- dt^2 +dx_1^2+ dx_2^2+ dx_3^2\right )
+ H^{1/2} \left ( dr^2 + r^2 d\Omega_5^2 \right ), \label{D3brane}
$$
where
$$H= 1+{R^4\over r^4}\ .
$$
The s-wave of a minimally coupled massless scalar satisfies
\beq
\left [\rho^{-5} {d\over d\rho} \rho^5 {d\over d\rho} +
1 + {(\omega R)^4\over \rho^4} \right ] \phi(\rho) =0\ , \label{Coul}
\eeq 
where $\rho = \omega r$. 
For small $\omega R$ the absorption cross section was obtained by
Klebanov \cite{Klebanov} by solving the matching problem of an
approximate solution in the inner region and the outer region. The
following result was  obtained,
\beq 
\sigma_{\rm 3-brane}= {\pi^4\over 8}\omega^3R^8 \ .
\label{three-brane}
\eeq
And this is also precisely given from 4d $\cal{N}$=4 SYM theory as
a three point vertex of the bulk field and the fields of SYM
theory which are given by the Born-Infeld action.

\vspace{.3cm} In this case, the 4d $\cal{N}$=4 SYM theory is
described by  string theory or by  supergravity near the horizon
of the D3-branes. In this limit, the geometry is approximated by
Ad$S_5\times S^5$ and the equation of motion for the dilaton
($\phi$) or for the graviton polarized parallel to the D3 brane is
written as \beq \left(\partial_{r}^2+{5\over
r}\partial_{r}+M_g^2[1+(R/r)^4]\right)\phi=0, \label{glueball}
\eeq where $M_g$ denotes the 4d mass of the dilaton. However, for
Ad$S_5\times S^5$, there is no finite $M_g$ which gives a
normalizable wave function $\phi$ since $\cal{N}$=4 SYM is not a
confining theory. To obtain a confining YM theory, the bulk
geometry should be deformed to an appropriate form. In this case,
however, the geometry has a complicated form especially in the
infrared region.

\vspace{.3cm} A simple and tractable approximation for such a
confining geometry can be given by introducing an infrared cutoff
at an appropriate point of the coordinate $r$, say $r=r_m$. In
this case, the geometry Ad$S_5\times S^5$ is used in the region
$r_m<r<\infty$. This approximation would be justified since the
main deformation of the geometry is in the infrared region,
$r<r_m$, and this region is cut off now. In fact, we can obtain a
discrete mass spectrum for the glueball by solving
(\ref{glueball}) and imposing the boundary condition
$\partial_r\phi |_{r_m}=0$. When we adopt the lattice result for
the glueball mass, $M_g\sim 1.5$ GeV, the IR cutoff is set at
$r_m\sim 0.4$ in
 units where  $R=1$. Here we used the data in \cite{ISM} since
the temperature dependence of the glueball mass is given in this
reference. We see that it is consistent with our analysis given in
 Fig.\ref{mg}.

\vspace{.3cm} Now we turn to the absorption cross section for the
deformed D3 branes, whose horizon limit background is given by the
Ad$S_5\times S^5$ with IR cutoff. Then the background, which
describes also off the horizon limit, can be written as
(\ref{D3brane}) with the IR cutoff $r_m$. So the matching problem
is solved in a similar way as that developed by Klebanov. The only
difference is the existence of the IR cutoff, and the equation is
restricted to the region $r_m<r<\infty$. We call the region near
$r_m$  the inner region.

\vspace{.5cm} While the wave equation is solved  in terms of
Bessel functions at small $\rho$ \cite{Klebanov}, we here solve
the equation at some small but finite $r$, namely at the IR cutoff
$r_m (<R)$. To study the wave function in this region,
Eq.(\ref{Coul}) is solved by rewriting it according to \cite{DGM}
as \beq \left [\partial_{\tau}^2+ \omega_m^2 \right ] \phi(\tau)
=0\ , \quad \tau=-{1\over 4\rho^4}\ , \label{Coul-m} \eeq
$$ \omega_m=\rho_m^5\sqrt{1+(\omega R)^4/\rho_m^4}, \quad \rho_m=\omega r_m$$
The solution of this form of equation can be matched to the outer solution.
Then, near $r_m$, the incoming wave is given as
\beq
 \phi(\tau)_{\rm ~in}=e^{-i\omega_m\tau} \label{small-2}
\eeq
and this is approximated for small $\tau$ as
\beq
 \phi(\tau)_{\rm ~in}=1-i\omega_m\tau\, \label{cutoff}
\eeq

\vspace{.3cm}
In the outer region,
(\ref{Coul}) is estimated by substituting
$\phi = \rho^{-5/2} \psi$. Then we get
\beq 
[{d^2\over d \rho^2} - {15\over 4 \rho^2}
+1 + {(\omega R)^4\over \rho^4}] \psi =0\ . \label{Coulthree}
\eeq 
Now the last term is negligible for $\rho \gg (\omega R)^2$,
where (\ref{Coulthree}) is solved in terms of Bessel functions as
\beq 
\phi_{\rm out} = A\rho^{-2}
\left [J_2(\rho)+b
i N_2(\rho)  \right ] \ , 
\label{outer}
\eeq 
 $J$ and $N$ being the Bessel and Neumann functions.

\vspace{.5cm}
From the matching of the solutions,
(\ref{small-2}) and (\ref{outer}), in the overlapping region of $r$,
we obtain $A=1$ and $b=0$. Then
\beq 
\phi_{\rm out} = {1\over 8}\rho^{-2}J_2(\rho)\, .
\label{outer-2}
\eeq 
In terms of the formula for the invariant flux $F$,
\beq
 F={1\over 2i}\sqrt{-g}~g^{~rr}(\phi^{*}\partial_r\phi-c.c.), \label{flux}
\eeq the fluxes of the incoming wave are obtained as \beq
 F_{\rm inner}=\omega g_m,
\eeq
$$ g_m=r_m^5\sqrt{1+(R/r_m)^4}
$$
 near $r_m$, and \beq F_{\rm outer}={32\over \pi}\omega^{-4}.
\label{Fout} \eeq for the outer region. Then the absorption
probability at $r_m$ is given as \beq
 {\cal{P}}=
|{F_{\rm inner}\over F_{\rm outer}}|={\pi\over 32}\omega^5 g_m\, . \label{P0}
\eeq
We notice that the power of $\omega$ is different from the case of D3
background without the IR cutoff.

\vspace{.3cm}
The absorption cross-section $\sigma$ is
related to the s-wave absorption probability by \cite{DGM}
$$ \sigma = {(2 \pi)^{5}\over \omega^{5} \Omega_{5}} {\cal P}
\ ,$$
where
$$\Omega_D = {2 \pi^{{D+1\over 2}}\over \Gamma \left (
{D+1\over 2}\right ) }
$$
is the volume of a unit $D$-dimensional sphere.
Thus, for the 3-brane we find \footnote{By absorption cross section
we will consistently mean
the cross section per unit longitudinal volume of the
brane.}
the following cross section at the infrared cutoff $r_m$,
\beq 
\sigma_{\rm IR}= {\pi^3g_m}\ .
\label{three}
\eeq 
We notice that this result is independent of $\omega$ contrary to the case
of the background without the cutoff,
and it is proportional to the area of the five dimensional sphere
with radius $r_m$. So the situation is similar to the absorption
by a black body in the six dimensional space.

{In the case of  background without the cutoff, the cross
section is explained by the decay amplitude of the incoming scalar
to two gluons or other fields included in the SYM theory
\cite{Klebanov,GKT,GubserKlebanov}. In this case, the
semiclassical calculation is useful because of the conformal
invariance of the YM theory. And the gluons are not bounded to
glueballs in spite of the interactions with a strong coupling
constant. In the present case, however, the conformal invariance
of the gauge theory is of course lost and the gluons are confined
and bounded to glueball states. Then we expect that the incoming
scalar of energy $\omega$, which is in the range $M_g < \omega <
2M_g$, will form a glueball after it coupled to the gluon pair.
For $\omega > 2M_g$, there appears the possibility of pair
production of glueballs, and it would be possible to calculate the
pair production cross section as in the YM case. But the
interaction between the glueballs is complicated and strong. It
would be difficult to get some useful results from the 4d field
theory side. This point will be discussed more in the future. }


\section{Extension to finite temperature}

\vspace{.5cm}
\noindent For the finite temperature gauge theory, the corresponding gravity
background solution is given by the
nonextremal black three-brane. It has the form
\cite{HorowitzStrominger,DuffLu}
\begin{eqnarray}
  ds^2 &=& H^{-1/2}(r) [-f(r)dt^2 + d{\vec{x}}~^2] \nonumber\\
  & & +H^{1/2}(r) [f^{-1}(r) dr^2 + r^2 d\Omega_5^2] \, ,
  \label{metric}
\end{eqnarray}
where as before $H(r)=1+R^4/r^4$ and $f(r)=1-r_0^4/r^4$. The
Hawking temperature of this metric is \beq
 T={r_0\over \pi R^2}\ ,
\eeq and the metric of the previous section is obtained for $T=0$.
{It is well known that  gravity under this background in the
near-horizon limit of D3 branes describes the high temperature
deconfinement phase of the YM theory. On the other hand, we have
proposed that the low temperature confinement phase could be
obtained by introducing the IR cutoff, $r_m$, in this holographic
model as $r_0 <r_m (<R)$ \cite{GY,GNY}. When the two phases are
separated at $r_0=r_m$,  the critical temperature $T_c$ is given
as \beq
 T_c={r_m\over \pi R^2}\sim 127 ~{\rm MeV} \ .
\eeq

\vspace{.3cm} Next we consider the $s$-wave radial equation for a
minimally coupled scalar (such as the graviton polarized parallel
to the brane),
$\partial_\mu(\sqrt{-g}g^{\mu\nu}\partial_\nu\phi)=0$. For the
metric (\ref{metric}) this equation acquires the form \beq
\partial_{\rho}(\rho^5f\partial_{\rho}\phi)+{H\rho^5\over f}\phi=0\ .
\label{radial1} \eeq We solve this equation in two phases,  one at
low and one at high temperature, and give the absorption cross
sections in each phase.}

\vspace{.3cm}
\subsection {Low temperature phase $T<T_c$}
By introducing the IR cutoff  $r_0 <r_m (<R)$, the low temperature
phase is obtained for the range of $r_m<r<\infty$ in terms of the
metric (\ref{metric}). Here, we consider $r_m$ to be independent
of the temperature or of $r_0$. We estimate the absorption cross
section at the IR cutoff as in the previous section. Near $r=r_m$,
Eq.(\ref{radial1}) is rewritten as \beq \left
[\partial_{\tilde{\tau}}^2+ {\omega}_m^2 \right ]
\phi(\tilde{\tau}) =0\ , \label{Coul-m2} \eeq \beq \quad
{\partial\tilde{\tau}\over\partial r}={1\over \rho^5f}\
,\label{Coul-m3} \eeq where we notice $
{\omega}_m=\rho_m^5\sqrt{1+(\omega R)^4/\rho_m^4}, \quad
\rho_m=\omega r_m.$ Then at the IR cutoff, the incoming wave is
given as \beq
 \phi(\tilde{\tau})_{\rm ~in}=e^{-i{\omega}_m\tilde{\tau}},
\label{cutoff-22}
\eeq
and this is approximated for small $\tilde{\tau}$ as
\beq
 \phi(\tilde{\tau})_{\rm ~in}=1-i{\omega}_m\tilde{\tau}.\, \label{cutoff-2}
\eeq

\vspace{.3cm} On the other hand, in the region of large $\rho$, we
can approximate  $H\sim 1$ and $f\sim 1$, then we arrive at the
same form of solution as in  the previous section,
$$ 
\phi = \tilde{A}\rho^{-2}
\left [J_2(\rho)+\tilde{b}
i N_2(\rho)  \right ] \ . 
$$ 
As in the previous section, the outer solution is given as
$$ 
\phi_{\rm out} = {1\over 8}\rho^{-2}J_2(\rho)\, ,
$$ 
and this matches with (\ref{cutoff-22}) at an appropriate region
near $r_m$. Then we arrive at the same result for the fluxes of
the incoming wave as that obtained for $T=0$, \beq
 F_{\rm inner}=\omega g_m, \label{Fin2}
\eeq
where $g_m=r_m^5\sqrt{1+(R/r_m)^4}$ as shown above.
We should notice here that the factor $f(\rho)$ has disappeared in the final
result of (\ref{Fin2}) due to the exact cancellation between the wave function
and the measure in (\ref{flux}). As a result, $F_{\rm inner}$ is independent of
the temperature.

\vspace{.3cm}
 Now, $F_{\rm outer}$ is the same as given in
(\ref{Fout}) at zero temperature. Then all other quantities are
the same as in the zero temperature case. The absorption
probability at $r_m$ is given as \beq
 {\cal{P}}=
|{F_{\rm inner}\over F_{\rm outer}}|={\pi\over 32}\omega^5 g_m\, ,
\eeq and we obtain the absorption cross section at low
temperature, \beq \sigma_{\rm 3-brane}^{\rm low T}= {\pi^3g_m}\,
.\label{three-ab}
\eeq 
This  result is seen to be independent of the temperature and it
has the same form as that obtained at zero temperature. {This
point can be understood as follows. Since $T < T_c << M_g$,  the
thermal fluctuations of glueballs would be suppressed because we
need large energy, at least $\omega > M_g$, to obtain a definite
thermal effect, which should  in principle be expressed by some
kind of temperature dependence, according to a thermal field
theory for the glueball gas. }

\vspace{.3cm} {In the above discussion we neglected the
temperature dependence of the glueball mass $M_g(T)$, but it will
be important.
It is estimated here by 
solving the equation of motion
of the dilaton near the D3 horizon limit of (\ref{metric}).}

\begin{figure}[htbp]
\vspace{.3cm}
\begin{center}
  \includegraphics[width=11.5cm]{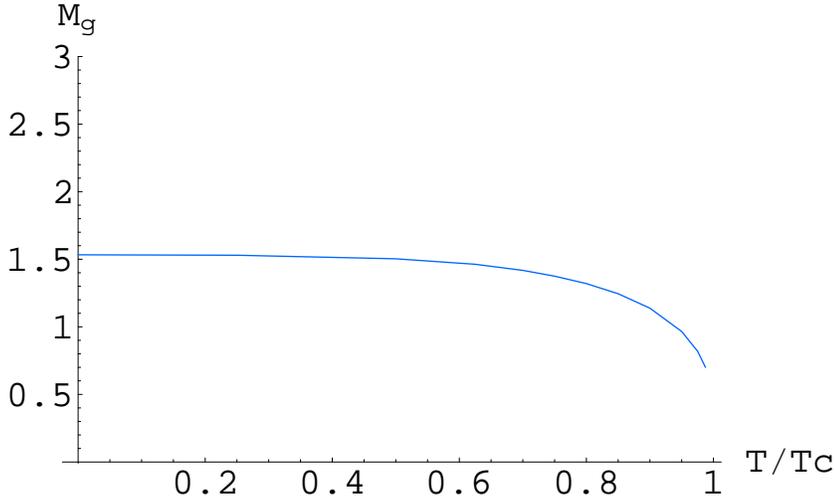}
\caption{Glueball mass vs temperature, for $r_m=0.4$ and $R=1$.
 \label{mg}}
\end{center}
\end{figure}
The equation to be solved is \beq
\left(\partial_{r}^2+\partial_{r}\left(\log(r^5
f(r))\right)\partial_{r} +M_g^2{1+(R/r)^4\over f^2}\right)\phi=0,
\eeq and the mass eigenvalue $M_g$ is obtained by finding the
normalizable solutions of this equation. The numerical results are
given in  Fig.\ref{mg} for $r_m=0.4$ and $R=1$. We observe its
typical $T$-dependence, $M_g\to 0$ for $T\to T_c$, and this
behavior can be understood by the last term in the equation. The
similar behaviour has been seen also in \cite{GY,GNY} and it is
consistent with the lattice analysis \cite{ISM} as pointed out
above. From these results we can say that the above
$T$-independence of (\ref{three-ab}) would be modified near $T_c$
since $M_g$ becomes small at this point and the probability of
pair production of glueballs can not be neglected in this region.

It is obvious that the temperature dependence of the mass comes
from the factor $f$ in the equation. On the other hand, the factor
$f$ is absent in (\ref{three-ab}). Therefore,  the temperature
dependence has to be included in the parameter $r_m$ itself, when
$\sigma_{\rm 3-brane}$ is temperature dependent. In other words,
$r_m$ will vary with  temperature, at least near $T=T_c$. This
point will be discussed  in the next sub-section.

\vspace{.5cm}
\subsection {High temperature phase}
At high temperature, $r_m<r_0$, we consider the region of $r_0<r<\infty$
where we can see the deconfinement phase of the gauge theory. Here we estimate
the absorption at the horizon $r_0$.
Near $r=r_0$, Eq.\
(\ref{radial1}) has the form
\begin{equation}
  \phi'' + {\phi'\over r-r_0} + \left({\omega\over 4\pi T}\right)^2
{\phi\over(r-r_0)^2}= 0 \, .
  \label{phihor}
\end{equation}
The matching problem has been solved as in \cite{PSS}, and we
obtain the absorption probability
$${\cal P}=|{F_{\rm near~ horizon}\over F_{\rm outer}}|=
   {\pi\over 32}\omega^5r_0^3R^2
$$
and the cross section
\beq
\sigma_{\rm 3-brane}^{\rm high T}= {\pi^3r_0^3R^2}\ .
\label{three-high}
\eeq 
Then the cross section increases like $T^3$ in the high
temperature QGP phase.

\vspace{.3cm}
On the other hand, the low temperature result Eq.(\ref{three-ab}) is rewritten
for the case $R/r_m >> 1$ as
\beq
 \sigma_{\rm 3-brane}^{\rm low T} 
\sim \pi^3r_m^3R^2\, , \label{three-ab-2} \eeq and we  see that
this is equivalent to (\ref{three-high}) for $r_0=r_m$. This
implies that the thermal activity is frozen by the value at
$T=T_c$ for $T<T_c$ since the glueball mass is much larger than
$T_c$. All the degrees of freedom contribute to the thermal
property for $T>T_c$ and we obtain the temperature dependence of
(\ref{three-high}).


{As for the IR cutoff parameter $r_m$, it has been introduced
by hand as a constant which is independent of $r_0$. However, it
may have a temperature dependence. In such a case,} we can refine
the description and write  $r_m=r_m(r_0)$,  imposing the following
condition \beq
 r_m(T_c)=r_c, \quad T_c={r_c\over \pi R^2}.
\eeq
 Then the value of $T_c$ is also changed to a larger value
than the 127 MeV given above.
{In fact, it is estimated as $T_c=270$
MeV in the lattice gauge theory \cite{ISM}. When we take this larger value,
we obtain $r_c\sim 2r_m(T=0)$. And we thus obtain the temperature
dependence of $\sigma_3$
\beq
\sigma_{\rm 3-brane}^{\rm low T}(T_c)=
         8\sigma_{\rm 3-brane}^{\rm low T}(0)\ .
\label{T-dep-sigma}
\eeq 
}
However, this analysis should be studied in more detail and it is outside the
scope of the present work.

\section {Shear viscosity}
We briefly review the
notion of viscosity in the context of finite-temperature field theory.
Consider a plasma slightly out of equilibrium, so that there is local
thermal equilibrium everywhere but the temperature and the mean
velocity slowly vary in space.  We define, at any point, the local
rest frame as the one where the three-momentum density vanishes:
$T_{0i}=0$.  The stress tensor, in this frame, is given by the
constitutive relation,
\begin{eqnarray}
  T_{ij} &=& \delta_{ij}p - \eta \biggl(\partial_i u_j + \partial_j u_i -
  {2\over3} \delta_{ij} \partial_k u_k\biggr)\nonumber\\
  & & - \zeta \delta_{ij} \partial_k u_k
  \, ,
\end{eqnarray}
where $u_i$ is the flow velocity, $p$ is the pressure, and $\eta$ and
$\zeta$ are, by definition, the shear and bulk viscosities
respectively.  In conformal field theories like the $\cal{N}$=4 SYM theory,
the energy momentum tensor is traceless, ${T^\mu}_\mu=0$, so
$\varepsilon\equiv T_{00}=3p$ and the bulk viscosity vanishes
identically, $\zeta=0$.

All kinetic coefficients can be expressed, through Kubo relations, as
the correlation functions of the corresponding currents \cite{Hosoya}.
For the shear viscosity, the correlator is that of the stress tensor,
\begin{eqnarray}
  \eta(\omega) &=& 
{1\over2\omega}\int\!dt\,dx\, e^{i\omega t}
   < [T_{xy}(t, x), T_{xy}(0, 0)] > \nonumber\\
  &=& 
  {1\over2\omega i}[G_{\rm A}(\omega)-G_{\rm R}(\omega)]\, ,
  \label{Kubo}
\end{eqnarray}
where the average $<\ldots >$ is taken in the equilibrium thermal
ensemble, and $G_{\rm A}$ and $G_{\rm R}$ are the advanced and
retarded Green functions of $T_{xy}$, respectively.  In Eq.\
(\ref{Kubo}), the Green functions are computed at zero spatial
momentum. Though Eq.\ (\ref{Kubo}) can, in principle, be used to
compute the viscosity in weakly coupled field theories, this
direct method is usually very cumbersome, since it requires
resummation of an infinite series of Feynman graphs.  This
calculation has been explicitly carried out only for scalar
theories \cite{Jeon}.  A more practical method is to use the
kinetic Boltzmann equation, which gives the same results as the
diagrammatic approach \cite{JeonYaffe}.



The static shear viscosity is defined in the limit of $\omega\to 0$
of the correlator of energy momentum tensor,
\begin{eqnarray}
  \eta(0) &=& \lim_{\omega\to0}{1\over2\omega}\int dt\,dx\, e^{i\omega t}
\langle [T_{xy}(t, x), T_{xy}(0, 0)] \rangle \nonumber\\
  &=& {1\over 2\kappa^2}\sigma(0) 
  \label{Kubo-2}
\end{eqnarray}
In the high temperature phase, the viscosity is given
from (\ref{three-high}) as \cite{PSS}
\beq
   \eta_{\rm H}={\pi N^2\over 8}T^3\, . \label{High-T}
\eeq
On the other hand, in the low temperature phase, we obtain by using
Eqs.(\ref{three-ab}) and (\ref{Kubo-2}),
\beq
 \eta_{\rm L}= {\pi N^2 \over 8}\left({r_m\over \pi R^2}\right)^3
    \left({r_m\over R}\right)^2\sqrt{1+\left({R\over r_m}\right)^4}\, \label{Low-T}.
\eeq When we do not consider $z_m$ as the T-dependent parameter,
this low temperature viscosity is independent of the temperature.
When we restrict ourselves to the case $r_m/R << 1$,  it is
approximated as \beq
   \eta_{\rm L}\sim {\pi N^2 \over 8}\left({r_m\over \pi R^2}\right)^3
    ={\pi N^2 \over 8}T_c^3 \, .
\eeq
It coinsides with (\ref{High-T}) at $T=T_c$.

\begin{figure}[htbp]
\vspace{.3cm}
\begin{center}
  \includegraphics[width=11.5cm]{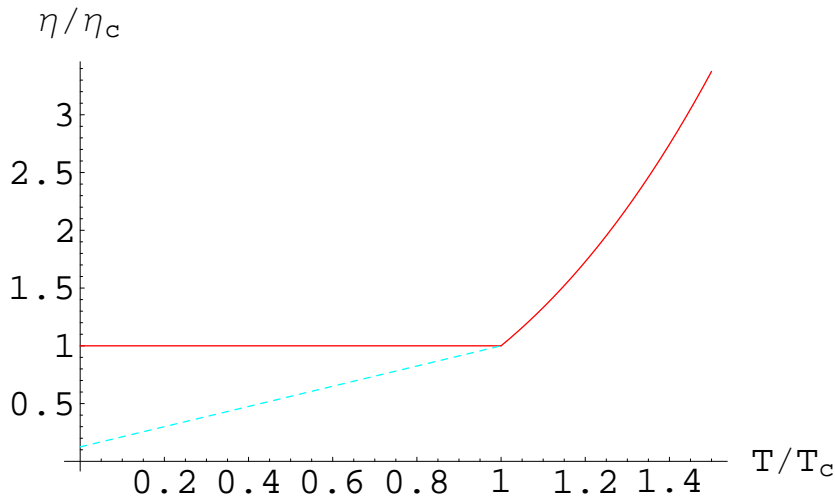}
\caption{$\eta/\eta_c$ is shown, where $\eta_c={\pi N^2 \over
8}T_c^3$. Temperature is scaled by $T_c$. The dashed line shows
the case of T-dependent $r_m$ with $T_c=270$ MeV.
 \label{viscosity}}
\end{center}
\end{figure}

{This result seems to be reasonable. Since the lowest state of
the glueball has a large mass ($\sim$ 1.5 GeV) and its gas would
not be thermodynamically active below $T_c\sim 127$ MeV,  we
expect its T-dependence to be negligible.
When we adopt the critical
temperature as $T_c\sim 270$ MeV \cite{ISM}, however, the cross section
changes as given in (\ref{T-dep-sigma}). As a result, we expect
the T-dependence of the viscosity for YM theory to be as shown in
 Fig.\ref{viscosity}}.

\section{Summary}
In this paper we have studied the absorption cross section and
viscosity of pure Yang-Mills theory at low temperature where the
theory is in the confinement phase. The gluons are confined and we
expect that the glueball states are the dominant modes in this
phase. In our holographic model, the infrared cutoff $r_m$ is
introduced as a parameter, and it is fixed by lowest mass of the
glueball which is estimated to be about $M_g=$ 1.5 GeV. Using this
$r_m$, the critical temperature of gluon confinement is estimated
to be $T_c\sim 127$ MeV.

We expect that the thermal properties of the YM theory below $T_c$ will be determined
 by the glueball gas. However, for $T<T_c$ the glueball mass is much larger than the
  temperature, then the glueball gas will not show any active temperature dependence.
   As expected, we find that both the absorption cross section and  the shear viscosity
   are independent of the temperature. The value of the
shear viscosity for $T<T_c$ is frozen at the value corresponding
to the critical point up to $T=0$. The only possible temperature
dependence of the viscosity will be seen through the temperature
dependence of the cutoff itself as $r_m=r_m(T)$. This is an
interesting point, to which we will return in the future. But it
is an open problem at the present stage.

We have not considered the light flavor quarks and the light
mesons constructed by them. If we take into account   those
mesons, the thermal properties at low temperature will be changed
due to the light flavored mesons \cite{chen06}. This point is also
an open problem at the moment.

Finally, let us make some  remarks from a fundamental viewpoint on
possible physical interpretations of
 the quantity $r_m$. From the above
formalism it follows that $r_m$ can be looked upon in two
different ways. First, $r_m$ can be considered to play the role of
an extra horizon, in addition to the conventional horizon $r_0$
that is present at finite temperature. While Policastro et al.
\cite{PSS}  got $\sigma(\omega) \sim \omega^3$ for small $\omega$
at $T=0$ and thus $\sigma(0)=0$, we got $\sigma_{\rm IR}$ to be
finite at the infrared cutoff (Eq.~(\ref{three})), at zero
temperature. This is to be compared with  the result
$\sigma(0)=\pi^3r_0^3R^2$ in \cite{PSS}, which is finite at finite
temperature. Secondly, our result for the shear viscosity at high
and low temperature is given by Eqs.(\ref{High-T}) and
(\ref{Low-T}). Here, the behaviour of $\eta_L$ is seen to be
influenced by $r_m$. It is of interest to compare also this result
with that obtained in \cite{PSS}: $\eta= f(g^2N)N^2T^3$, where
$f(x)$ is a function satisfying $f(x) \sim x^{-2}\ln^{-1}(1/x)$
when $x \ll 1$ and $f(x)=\pi/8$ when $x \gg 1$. We thus see that
the quantity $ r_m$ can manifest itself in different ways
physically. It hardly seems possible to associate $r_m$ simply
with a unique conventional physical quantity.

\section*{Acknowledgments}

The authors are very grateful to M. Tachibana
for useful discussions and comments throughout this work.
This work has been supported in part by the Grants-in-Aid for
Scientific Research (13135223)
of the Ministry of Education, Science, Sports, and Culture of Japan.


\newpage

\end{document}